\documentstyle[12pt]{article}
\def\beq{\begin{equation}}
\def\eeq{\end{equation}}
\def\bea{\begin{eqnarray}}
\def\eea{\end{eqnarray}}
\def\bq{\begin{quote}}
\def\eq{\end{quote}}

\def\gappeq{\mathrel{\rlap {\raise.5ex\hbox{$>$}}
{\lower.5ex\hbox{$\sim$}}}}

\def\lappeq{\mathrel{\rlap{\raise.5ex\hbox{$<$}}
{\lower.5ex\hbox{$\sim$}}}}

\begin{document}
\topmargin -0.5cm
\oddsidemargin -0.8cm
\evensidemargin -0.8cm
\pagestyle{empty}
\begin{flushright}
{CERN-TH.6946/93}\\
{CERN/LAA/93-23}
\end{flushright}
\vspace*{5mm}
\begin{center}
{\bf A SEARCH FOR EXACT SUPERSTRING VACUA} \\
\vspace*{1cm}
{\bf A. Petermann } \\
\vspace{0.3cm}
TH Division, CERN,
CH - 1211 Geneva 23 \\
\vspace*{0.3cm}
and \\
\vspace*{0,3cm}
{\bf A. Zichichi}\\
\vspace{0.3cm}
PPE Division, CERN, CH - 1211 Geneva 23 \\
\vspace*{2cm}
{\bf ABSTRACT} \\ \end{center}
\vspace*{5mm}
\noindent
We investigate $2d$ sigma-models with a $2+N$ dimensional Minkowski
signature target space metric and Killing symmetry, specifically
supersymmetrized, and see under which conditions they might lead to
corresponding exact string vacua. It appears that the issue relies
heavily
on the properties of the vector $M_{\mu}$, a reparametrization term,
which needs to
possess a definite form for the Weyl invariance to be satisfied. We
give, in
the $n = 1$ supersymmetric case, two non-renormalization theorems from
which
we can relate the $u$ component of $M_{\mu}$ to the $\beta^G_{uu}$
function. We work
out this $(u,u)$ component of the $\beta^G$ function and find a
non-vanishing
contribution at four loops. Therefore, it turns out that at order
$\alpha^{\prime
4}$, there are in general non-vanishing contributions to $M_u$ that
prevent us from
deducing superstring vacua in closed form.

\vspace*{3cm}


\begin{flushleft} CERN-TH.6946/93 \\
CERN/LAA/93-23 \\
July 1993
\end{flushleft}
\vfill\eject

\setcounter{page}{1}
\pagestyle{plain}
\section{Introduction}

To find exact string vacua is one of the most interesting problems
of string theory. In fact, the symmetries of the vacuum are the
symmetries of the
World \cite{aaa}, the constants of the vacuum are the constants of the
World
\cite{bb} and therefore to know the string vacua is indeed the first
crucial step
towards a TOE.

In the past few months, attention has been brought to $2d$ finite
sigma-models with a $2+N$ dimensional Minkowski signature target space
metric with a
covariantly constant null Killing vector. They can be considered as
describing string
tree-level backgrounds consisting in plane gravitational wave-type,
supplemented
by a dilaton background. Such kinds of solution of Einstein equations
have been
discussed long ago by Brinkmann \cite{cc}.

These models have been extensively studied
\cite{dd} in the bosonic case and show several important features such
as:
\begin{description}
\item{(1)}	the
UV-finiteness (on shell) of these models;
\item{(2)}	that, given a non-conformal $\sigma$-model
with Euclidean $N$ dimensional target space (the so-called transverse
space), there
exists a conformal invariant Minkowskian $\sigma$-model in $2+N$
dimensions;
\item{(3)}	that,
because of the Killing symmetry, the $2+N$ dimensional metric does not
depend on one
of the two extra coordinates $u$ and $v$;
\item{(4)}	the fact that the $2+N$ dimensional metric
is expressed in terms of the running coupling of the transverse theory.
\end{description}
Later,
Tseytlin \cite{ee} and the present authors \cite{ff} discussed
independently a
specific supersymmetric extension of this class of models, which allows
one to know
exactly the transverse beta function, thanks to a non-renormalization
theorem
\cite{ggg}, and therefore to be able to give the line element in closed
form.

But, having in mind exact string vacua identification, we need also to
know an appropriate dilaton field such that the Weyl invariance
conditions are
satisfied. Only then will the resulting models correspond to string
vacua. That such
a dilaton field exists has been proved order by order in a perturbation
expansion in
$\alpha^{\prime}$. However, this does not allow in general a formulation
in closed
form. The crucial point is a piece of information on $W_u$, the
$u$-component of
$W_{\mu}$, a
vector that enters linearly in the reparametrization vector $M_{\mu}$
(cf. section 2);
$W_{\mu}$ is a covariant vector originating through the mixing under
renormalization
of dimension 2 composite operators \cite{hh}. While earlier discussions
on this point
had been based on conjectures (see \cite{ee}, \cite{ff}, \cite{jj},
\cite{kk}), none
of them seems to be satisfactory, being either incomplete \cite{ee},
\cite{kk} or only
necessary but not sufficient \cite{ff} (see also \cite{lll}).

It is the aim of this
paper to go deeper into this question. In order to get concrete
information, we have
been looking at the first non-vanishing contribution to $W_u$, if it
exists at all.
In the bosonic case, it is well known \cite{dd} that $W_u$ starts to be
non-vanishing
at $\alpha^{\prime 3}$ and behaves like $u^{-3}$ ($u$ is a  light-cone
coordinate)
\footnote{For instance, the three-loop contribution to $\beta_{uu}$ due
to the
term $D_u R_{\alpha\beta\gamma\delta}D_uR^{\alpha\beta\gamma\delta}$
reads
$-\frac{\alpha^{\prime 3}}{16} \dot f^2 f^{-4}R_{ijk\ell}R^{ijk\ell}; i,
j, k,
\ell$  are transverse indices, and $f = bu$.}. In
the $n = 1$ supersymmetric case under examination in the present paper,
$W_u$ has been
computed to order $\alpha^{\prime 4}$ in two essentially different ways:
first
directly; and secondly via the non-renormalization theorem, which binds
$\partial_uW_u\equiv\dot W_u$ and the $\beta^G_{uu}$ component of
$\beta^G_{\mu\nu}$,
the beta function of $G_{\mu\nu}$. The two coinciding results show that
$W_u$, with a
$u^{-4}$ behaviour, starts to be non-vanishing at order $\alpha^{\prime
4}$ for general
$N$.

This work is organized as follows: in section 2 we recall the general
features of the $n = 1$ supersymmetric model, which we shall investigate
and the two
non-renormalization theorems uncovered in this model, in addition to the
well-known
theorems on the homogeneous K\"ahler transverse space (hence $n = 2$
supersymmetric). In
section 3, we give the results for the $\beta^G_{uu}$ component as
derived from two
different sources and for the direct computation of the $W_u$ component
of $W_{\mu}$. In
section 4, we give and discuss these results for a simple homogeneous
manifold as
transverse space. Finally, in section 5, we present our conclusions.

\section{The supersymmetric model and the non-renormaliza\-tion
theorems}

In the class of finite 2d $\sigma$-models introduced in Ref. \cite{dd},
the
$N$-dimensional  transverse space was supposed to be a Euclidean
symmetric space. Order
by order it was shown that there exists a dilaton which, together with
the $2+N$
dimensional metric background, solves the Weyl invariance conditions.
The metric of the
$2+N$ dimensional space turns out to give the line element
\bea
ds^2 =	G_{\mu\nu} dx^{\mu} dx^{\nu} = -2 du ~dv + f(u) \gamma_{ij} dx^i
dx^j	\nonumber
\\
	\mu , \nu = 0, \ldots, N+1; ~~i, j = 1, \ldots, N
\label{1}
\eea

With a specific choice of $f(u)$, the model is shown to be UV-finite;
$f(u)$ is
bound to satisfy a first-order RG equation
\beq
p\dot f = \beta(f) ;~~ \dot f = \partial_uf ;~~ p = {\rm constant}
\label{2}
\eeq
with $\beta(f)$ defined by the transverse $\beta^G_{ij}$
\beq
\beta^G_{ij} = \beta(f) \gamma_{ij}~.
\label{3}
\eeq

In order to complete the proof of finiteness of the model on a flat 2d
background,
the beta function of the $2+N \sigma$-model with target space metric
$G_{\mu\nu}$ has to
vanish up to a reparametrization term \cite{hh}:
\beq
\beta^G_{\mu\nu} + 2D_{(\mu} M_{\nu )}  = 0
\label{4}
\eeq
$M_{\mu}$ is not arbitrary and to establish that the $\sigma$-model
based on (1) is
Weyl-invariant, one needs to show that a dilaton field $\phi$ exists
such that
$M_{\nu}$ in (4) can be represented by
\beq
M_{\mu} = \alpha^{\prime}\partial_{\mu}\phi + \frac{1}{2} W_{\mu}
\label{5}
\eeq
the origin of $W_{\mu}$ having been specified in the Introduction.

The $n = 1$ supersymmetric extension (\cite{ee}, \cite{ff}) of the model
with bosonic
action \cite{dd} has been done as schematically indicated below.

One replaces the bosonic action
$$
I_b = (4\pi\alpha^{\prime})^{-1} \int d^2z \sqrt{g} [G_{\mu\nu}(x)
\partial_{\alpha}x^{\mu}\partial^{\alpha}x^{\nu} +
\alpha^{\prime}R^{(2)}\phi (x)]
$$
by the following superfield action
\beq
I = I_G + I_{\phi}
\label{6}
\eeq
with
\beq
I_G = (4\pi\alpha^{\prime})^{-1} \int d^2zd^2\theta
G_{\mu\nu}(X)DX^{\mu}\bar DX^{\nu}
\label{7}
\eeq
\beq
I_{\phi} = (4\pi )^{-1} \int d^2zd^2\theta R^{(2)}E^{-1}\phi (X)
\label{8}
\eeq
$D = \partial / \partial\bar\theta + \bar\theta\gamma^a\partial_a$
and $E^{-1}$ is the
determinant of the $n = 1$ super-vielbein.

It is then obvious how to specialize the
general metric $G_{\mu\nu}$ to a null Killing vector metric as in (1),
in terms of the
real superfields U, V and $X^i$.
\beq
I = (4\pi\alpha^{\prime})^{-1} \int d^2zd^2\theta [-2DU\bar DV + g_{ij}
(U,X) \cdot
DX^i\bar DX^j]~.
\label{9}
\eeq

The generalization of the bosonic case studied in \cite{dd} to the
supersymmetric case
is then straightforward and the finiteness condition for symmetric
spaces (2)
\footnote{With metric tensor given by $g_{ij}(u,x) =
f(u)\gamma_{ij}(x)$.} will also be
determined by the beta-function of the transverse part of (9). If this
transverse space
is chosen to be K\"ahlerian, the $N$-dimensional part is $n = 2$
supersymmetric. The
choice in \cite{ee} and \cite{ff} was even more restrictive, assuming
the transverse
space to be a symmetric $\underline{\rm homogeneous}$ K\"ahler manifold.
This was
dictated by the fact that known examples of these manifolds exist, for
which the
beta-function reduces to its one-loop expression and is therefore
exactly known
\cite{mm}, \cite{ggg}. In this case, $\beta (f)$ reduces to a constant
$a$, depending
on the manifold chosen and its symmetries (remember that $f$ is the
inverse of the
generic transverse $\sigma$-model coupling $\lambda$); $f(u)$ becomes
equal to
\beq
f(u) = bu
\label{10}
\eeq
with $b = ap^{-1}$, and the transverse metric $g_{ij}(u,x)$ is
$bu\gamma_{ij}$.
Evidently the $n = 2$ supersymmetry is not shared by the full $2+N$
model with Minkowski
signature studied here and has only $n = 1$ by construction [cf. eqs.
(7), (9)].
However, use can be made of the result \cite{nn} that in the $n = 2$
case, the dilaton
coupling does not get renormalized, so that some quantities appearing in
the Weyl
anomaly coefficients of the $\underline{\rm transverse}$ part do vanish
in the minimal
subtraction scheme we use throughout (see \cite{ee} for details).
Therefore the $2+N$
model with $n = 1$ supersymmetry and with homogeneous symmetrical
K\"ahler transverse
subspace has a simplified structure, as compared with the generic $n =
1$
$\sigma$-models. However, the $u$ components of the various key
quantities such as
$W_u, \beta^G_{uu}$ and the ``interaction" part $\Phi (u)$ of the
dilaton field $\phi$,
expressed as \beq
\phi(u,v) = pv + qu + \Phi(u)
\label{11}
\eeq
for symmetric transverse spaces, do not seem to benefit from the special
properties of
the transverse part.

Nevertheless, one can formulate two non-renormalization theorems
concerning
the three quantities $\Phi(u), W_u$ and $\beta^G_{uu}$:
\begin{description}
\item{(1)}
$4\dot\Phi + W_u = 0$ beyond one loop,
\item{(2)}
$\dot W_u + 2\beta^G_{uu} = 0$ beyond one loop,
\end{description}
(at the one-loop level, however, $4\dot\Phi + W_u = N/2u$ and $\dot W_u
+ 2\beta^G_{uu}
= - N/2u^2)$.

As already emphasized in the Introduction, the issue of whether $W_u$
is identically zero or starts being non-vanishing at some high order is
crucial. For,
if zero, then the dilaton field can be integrated to a closed form. If
not, this closed
form will only be the starting value of a perturbation expansion in
$\alpha^{\prime}$
and the associated string vacuum, though existing, can only be expressed
perturbatively. No string vacuum can be given in closed form if $W_u
\not= 0$,
according to the present methods.

In a first attempt to clarify the situation, we computed both $W_u$ and
$\beta^G_{uu}$
in the $n = 1$ supersymmetric case. The results are shortly presented in
the next
section.

\section{$W_u$ and $\beta^G_{uu}$ at four loops}

We have explicitly worked out at four-loops the quantity $W_u$ and the
$(u,u)$ component
of the $n = 1$ supersymmetric beta-function in the simplest non-trivial
case, when the
subspace ($N$-dimensional transverse space) is locally symmetric.

The direct calculation
of $W_u$ comes out with the result
\beq
W_u = - \frac{\zeta (3)}{3(4\pi )^4}~ \dot ff^{-4}\cdot T_1~;
\label{12}
\eeq
with $T_1 = (R^{\phantom{a}[bc]}_{a\phantom{[bc]}d} +
R^{\phantom{a}bc}_{a\phantom{bc}d} \cdot
R^{d\phantom{e}a}_{\phantom{d}e\phantom{a}f}\cdot
R^{e\phantom{bc}f}_{\phantom{e}bc\phantom{f}}$.

Although we can derive the $\beta^G_{uu}$ from (12) and the
non-renormalization
theorem, we evaluated it, as a cross-check, from two different sources:
\begin{description}
\item{(1)}	the direct calculation \cite{oo} of the beta-function in $n =
1$
supersymmetric
non-linear generic $\sigma$ models, with the result
\beq
\beta^G_{uu} = - \frac{\zeta (3)}{3(4\pi )^4}   \dot f^2f^{-5}\cdot
T_2~;
\footnote{The covariant derivatives $D_u$ differ from ordinary
$\partial_u$, the
connections $\Gamma^i_{ju}$ being non-vanishing for $i, j$ transverse.}
\label{13}
\eeq
with $T_2 = R^{\phantom{a}bc}_{a\phantom{bc}d} \cdot
R^{d\phantom{ef}fa}_{\phantom{d}ef\phantom{a}} \cdot
R_{b\phantom{(ef)}c}^{\phantom{b}(ef)} + \frac{3}{4} R_{bcda}\cdot
R^{dae}_{\phantom{dae}f}\cdot R^{bc\phantom{e}f}_{\phantom{bc}e}$. ~~~
$R_{ijk\ell}$  means
$R_{ijk\ell} (\gamma )$;  \item{(b)} the simple derivatives of a
scalar built out from the sum of two different contractions of the
product of three
Riemann tensors \cite{lll}, confirming eq. (13). \end{description}

Accessorily, it can be verified that
\beq
\ddot\Phi = \frac{1}{2} \beta^G_{uu} = - \frac{1}{4} \dot W_u, ~{\rm
at~four~loops}
\label{14}
\eeq
so that
\beq
\beta^G_{uu} + \dot W_u + 2\ddot\Phi = \bar\beta^G_{uu} = 0~,
\label{15}
\eeq
which is the Weyl invariance condition for the $(u,u)$ component of the
gravitational
$\bar\beta^G_{\mu\nu}$ Weyl anomaly coefficient. This shows, if
necessary, that the
dilaton field $\phi$, eq. (11), will receive contributions from higher
order, which are
due to $\sigma$-model interactions.

\section{Concrete example}

We want to apply here the general formulas we have found in the previous
sections and verify accessorily that the non-renormalization theorems
are satisfied.
There are general Riemannian manifolds called spaces of constant
curvature, i.e. whose
curvature is independent both of the surface direction and the position
(for details,
see e.g. Ref. [3b], section 18). These manifolds have a particularly
simple curvature
tensor, said maximally symmetric, which reads
\beq
R_{hijk} = \frac{R}{N(N-1)}~(\gamma_{hj}\gamma_{ik} -
\gamma_{hk}\gamma_{ij})~,
\label{16}
\eeq
$R$ being the constant curvature and $N$ the dimension of space
(transverse in our
case).

Of course these manifolds, for instance the $N$-dimensional sphere $S^N$
embedded in
a Euclidean $R_{N+1}$ space, are Riemannian but not K\"ahlerian in
general. Therefore
they are not suitable for our transverse space, which we assumed to be
homogeneous
K\"ahler. However, it happens that for $N|=|2$ this Riemannian space is
also homogeneous
K\"ahler, due to the accidental isomorphism between $S^2 = SO(3)/SO(2)$
and $CP^1$. So,
for $N|=|2$, this model is homogeneous K\"ahler and possesses all the
properties wanted
for our $N$ -dimensional transverse space. Also the metric (16) can be
used and it is
straightforward to establish that (12) takes the value
\beq
W_u = \frac{\zeta (3)}{6(4\pi )^4}~\dot ff^{-4}\cdot R^3
\label{17}
\eeq
and (13)
\beq
\beta^G_{uu} = \frac{\zeta (3)}{3(4\pi )^4}~\dot f^2f^{-5}\cdot R^3
\label{18}
\eeq

Equations (17) and (18) obviously verify (14) and the
non-renormalization theorem of
section 2.

One notes also that the generic four-loop term contributes, for $n = 1$
supersymmetry, a quartic expression in terms of the curvature tensor,
which is the only
one to survive in locally symmetric spaces for general $N$. However,
using the metric
(16), it gives a contribution proportional to $(N-2)$ and therefore
vanishes in our
example for which $N = 2$. This exemplifies the consistency of the
approach: the quartic
term has a pure Riemannian origin, but must be absent in a K\"ahler
geometry, as it does
in our example.

\section{Conclusions}

In spite of the well-known fact that, even in the simplest models
with $n = 1$ supersymmetry, the four-loop beta-function is
non-vanishing, witnessing the
tree-level string theory graviton scattering modification to Einstein
action \cite{qq},
the model introduced by Tseytlin \cite{dd} and conveniently
supersymmetrized
\cite{ee}, \cite{ff} is so specific that the hope was not unreasonable
to see it
escaping the four-loop contributions to the $(u,u)$ component of the
beta-function. As a
matter of fact, and as a posteriori justification of our attempt, the
genuine
Riemannian $\alpha^{\prime 4}$ contribution to $\beta^G_{\mu\nu}$ ,
proportional to
$R^4$, produces
zero contribution to $\beta^G_{uu}$ in the present model \footnote{As it
does also
of course to $\beta^G_{ij}$ or $\beta^G_{iu}$ since the $N$-space has
been assumed
(homogeneous) K\"ahlerian.}, while it is fully contributing to the
generic $n = 1$
supersymmetric $\sigma$-model (see for instance \cite{lll}). This is an
example of its
specificity.

Summarizing, it turns out that only an $n = 4$ supersymmetry allows a
perfect
knowledge of the backgrounds in closed form. With a hyper-K\"ahler
transverse space, the
beta-function of this transverse part is identically zero, in all
renormalization
schemes (RS) and $f$ is constant. The metric is trivially simple.
Moreover, if
non-renormalization theorems might exist too, they will not necessarily
be the same as
those of section 2. But if $W_u$ starts getting non-zero contributions
at some high
order, even if it is at the four-loop level, there will necessarily be a
scheme in
which it vanishes identically and for which the beta function is still
identically
zero, since this property is RG-invariant. This is in contrast with the
situation in
the present paper. We can indeed find a scheme in which $W_u$ is zero at
all orders.
However, we will lose the exact knowledge of the metric due to $\beta
(f) = a$, the
``vanishing of all loop contributions but the first" being
$\underline{\rm not}$ an
RG-invariant property, but specific to particular schemes. We plan to
come back on
these general questions in a forthcoming publication \cite{mm}.

\vspace{0.3cm}
\noindent
{\bf Acknowledgements}

We would like to acknowledge several enlightening conversations and
correspondence with Arkady Tseytlin. One of us (AP) thanks also the
Theoretical Physics
Centre of the CNRS at Luminy-Marseilles for the hospitality.
\vfill\eject

\end{document}